\documentclass[aip,amsmath,twocolumn,amssymb,reprint]{revtex4-1} 

\usepackage[]{graphicx}
\usepackage{hyperref}
\hypersetup{
    pdfauthor={C. C. Lo},     
    pdfcreator={C. C. Lo},   
    pdfnewwindow=true,      
    colorlinks=true,       
    linkcolor=red,          
    citecolor=green,        
    filecolor=magenta,      
    urlcolor=cyan           
}

\begin{document}
\title{Stark shift and field ionization of arsenic donors in  $^{28}$Si-SOI structures}
\author{C. C. Lo}
\email{cheuk.lo@ucl.ac.uk.}
\affiliation{London Centre for Nanotechnology, University College London, London WC1H 0AH, U.K.}
\affiliation{Department of Electronic and Electrical Engineering, University College London, London WC1E 7JE, U.K.}
\author{S. Simmons}
\affiliation{Department of Materials, University of Oxford, Oxford OX1 3PH, U.K.}
\author{R. Lo Nardo}
\affiliation{London Centre for Nanotechnology, University College London, London WC1H 0AH, U.K.}
\affiliation{Department of Materials, University of Oxford, Oxford OX1 3PH, U.K.}
\author{C. D. Weis}
\affiliation{Accelerator and Fusion Research Division, Lawrence Berkeley National Laboratory, Berkeley, California 94720, U.S.A.}
\author{A. M. Tyryshkin}
\affiliation{Department of Electrical Engineering, Princeton University, Princeton, New Jersey 08544, U.S.A.}
\author{J. Meijer}
\altaffiliation{Present address: Division of Nuclear Solid State Physics, University of Leipzig, Leipzig, Germany}
\author{D. Rogalla}
\affiliation{RUBION, Ruhr-Universitaet Bochum, Germany}
\author{S. A. Lyon}
\affiliation{Department of Electrical Engineering, Princeton University, Princeton, New Jersey 08544, U.S.A.}
\author{J. Bokor}
\affiliation{Department of Electrical Engineering and Computer Sciences, University of California, Berkeley, California 94720, U.S.A.}
\author{T. Schenkel}
\affiliation{Accelerator and Fusion Research Division, Lawrence Berkeley National Laboratory, Berkeley, California 94720, U.S.A.}
\author{J. J. L. Morton}
\email{jjl.morton@ucl.ac.uk.}
\affiliation{London Centre for Nanotechnology, University College London, London WC1H 0AH, U.K.}
\affiliation{Department of Electronic and Electrical Engineering, University College London, London WC1E 7JE, U.K.}

\date{\today}

\begin{abstract}
We develop an efficient back gate for silicon-on-insulator (SOI) devices operating at cryogenic temperatures, and measure the quadratic hyperfine Stark shift parameter of arsenic donors in isotopically purified $^{28}$Si-SOI layers using such structures. The back gate is implemented using MeV ion implantation through the SOI layer forming a metallic electrode in the handle wafer, enabling large and uniform electric fields up to $\sim$\:2\:V/$\mu$m to be applied across the SOI layer. Utilizing this structure we measure the Stark shift parameters of arsenic donors embedded in the  $^{28}$Si SOI layer and find a contact hyperfine Stark parameter of  $\eta_a=-1.9\pm0.2\times10^{-3}\:\mu$m$^2$/V$^2$. We also demonstrate electric-field driven dopant ionization in the SOI device layer, measured by electron spin resonance. 
\end{abstract}

\maketitle 

Shallow donors in silicon are promising candidates for spin-based quantum information processing~\cite{kane98}. Although it has been known for decades that the coherence times of donor spins are relatively long at cryogenic temperatures~\cite{gordon58, feher59, castner67},  recent measurements with improved instrumentation, better understanding of the decoherence mechanisms and the availability of ultra high purity isotopically enriched nuclear spin-free $^{28}$Si crystals have led to dramatic improvements of the measured spin coherence times, exceeding seconds for phosphorus and bismuth donor electron spins~\cite{tyryshkin12, wolfowicz13}. Furthermore, the nuclear spin coherence times of ionized phosphorus donors have been found to exceed hours at cryogenic temperatures and 39 minutes at room temperature~\cite{saeedi13}. In addition to possessing extraordinary long spin coherence times, both the donor electron and nuclear spins can be read-out in a single shot~\cite{morello10, pla12, pla13} at the single-atom level.

A number of challenges remain in building on these advances to realize a scalable donor-based quantum computing architecture. Some such architectures require local addressability of donors under global microwave excitation, achieved, for example, by using electric fields  to tune donor spins in and out of resonance using the Stark shift~\cite{kane98}. As well as being a prerequisite for assessing the feasibility of such approaches, precise measurements of the donor Stark shift are also critical to aid our understanding and ability to model the donor energy levels accurately when embedded in nano-scale devices where large electric fields are usually present~\cite{mohiyaddin13}. In addition, measurements of ionized donor nuclear spins to date have relied on optical or thermal ionization processes~\cite{dreher12, saeedi13}. It is therefore desirable to develop a method, applicable to an ensemble, which achieves more controllable donor ionization and subsequent neutralization. 

The first measurement of the Stark shift parameters for shallow donors was reported using electron spin resonance (ESR) of $^{121}$Sb donors residing between interdigitated metallic electrodes evaporated on the silicon surface~\cite{bradbury06}. The geometry resulted in a non-uniform distribution of the electric field for the donor ensemble. In addition, the maximum electric field that could be applied was below 1\:V/$\mu$m, limited by leakage currents between the electrodes in high-field regions.  A parallel-plate capacitor-like structure in principle allows the application of uniform electric fields across the silicon and reduces the probability of having leakage currents. However, the voltage required to bias across the silicon substrate would need to be in the order of 100\:V to obtain appreciable Stark shifts for typical wafer thicknesses of 500\:$\mu$m. 

In this work we develop devices for ensemble measurements utilizing silicon-on-insulator (SOI) substrates, where the donors of interest are embedded in the thin SOI layer. We implement a novel back gate using high energy ion implantation into the handle wafer, which allows the efficient application of uniform and large electric fields across the SOI layer for both Stark shift measurements and controlled ionization of the dopants. This provides a platform for ensemble spin resonance measurements on electrically ionized and re-neutralized donor nuclear spins.

\begin{figure}
	\centering
	\includegraphics[width=8.0cm]{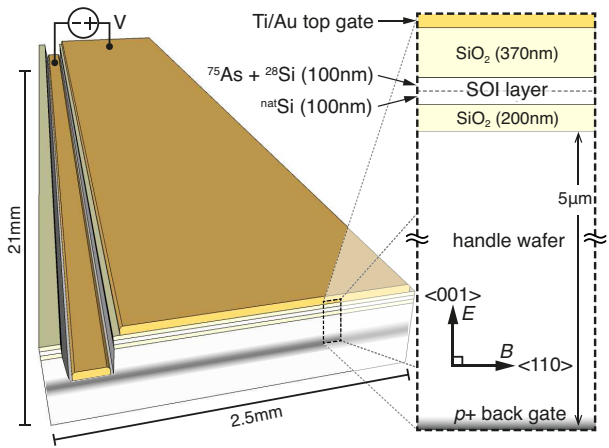}
	\caption{\label{f:1}{Schematic of the  $^{28}$Si-SOI device with a $p+$ back gate formed by MeV ion implantation. Inset: Cross-section of the device. 90\% of the $^{75}$As donors reside in the epitaxially grown  $^{28}$Si region of the SOI layer.}}
\end{figure}

The SOI substrates consisted of 100\:nm thick $p$-doped ($\sim$\:10\:$\Omega$cm) natural silicon ($^{\rm nat}$Si) with $\langle$100$\rangle$ orientation (SOITEC), and a similarly $p$-doped ($\sim$\:15\:$\Omega$cm) handle wafer underneath which normally freezes out at cryogenic temperatures. 100\:nm thick isotopically purified 99.92\% 28-silicon ($^{28}$Si) was epitaxially grown on the natural silicon SOI wafer (Lawrence Semiconductors), forming the  $^{28}$Si-SOI device layer. In order to create an effective back-gate for realizing a parallel plate geometry, a high energy boron (B) implantation of 4\:MeV at a dose of $2.2\times10^{15}$\:/cm$^2$ was performed, with almost all dopants implanted through the device and the 200\:nm thick buried oxide (BOX) layers, peaking at 5\:$\mu$m down from the handle wafer-buried oxide interface. The peak B concentration of $3\times10^{19}$\:/cm$^3$ is above the metal-insulator-transition, ensuring that the $p+$ region in the handle wafers remains metallic even at cryogenic temperatures. The added B concentration in the SOI layer resulting from the high energy implant is expected to be less than $1\times10^{15}$\:/cm$^3$ from device process simulations, and the lattice damage in the SOI layer is expected to be minimal due to ion channeling and most lattice damage is expected to be incurred at end-of-range~\cite{strm}. A 10\:nm silicon dioxide was subsequently grown on the SOI layer by dry oxidation at 900$^\circ$C. This oxidation step activated the implanted B and also reduced lattice damage in the handle wafer induced by the MeV implants.

\begin{figure*}
	\centering
	\includegraphics[width=16cm]{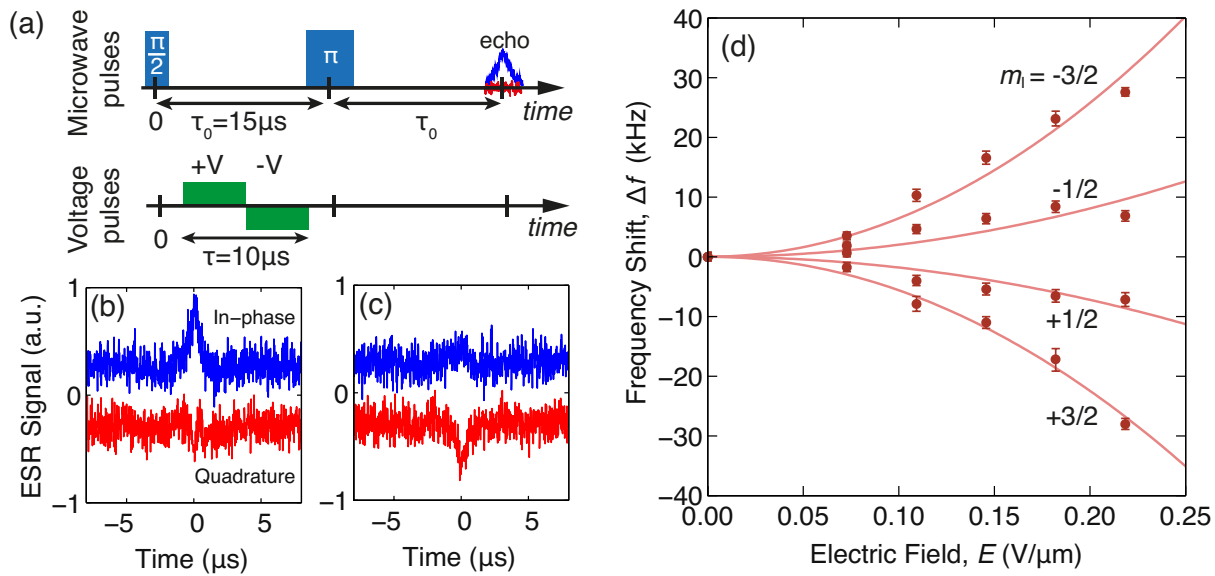}
	\caption{\label{f:2}{(a) Synchronization of the microwave and voltage (electric field) pulses used in the Stark shift experiments. Hahn echo signal for $m_I$=\:+3/2 of $^{75}$As donors centered at $2\tau_0$: (b) with no applied electric field, and (c) with an applied electric field of $E=\:0.23\:$V/$\mu$m. The phase shift from in-phase to quadrature is due to the Stark effect, with offset for clarity. (d) $^{75}$As donor frequency shifts for the four hyperfine-split lines. Solid lines represent fits to data with Eq\:\ref{e:1}.}}
\end{figure*}

The $^{28}$Si-SOI layer was then implanted with arsenic dopants ($^{75}$As) at 110\:keV at a dose of $4\times10^{11}$\:/cm$^2$ with 7$^\circ$ tilt. Arsenic was chosen to minimize dopant diffusion from the  $^{28}$Si layer to the natural silicon layer during thermal activation. In addition, the As nuclear spin of $I$=\:3/2 allows the clear elucidation of the contribution from the contact hyperfine induced Stark shift. The resulting peak donor concentration is located approximately 60\:nm away from the dry oxide-SOI layer interface.  360\:nm of low temperature oxide was subsequently deposited at 450$^\circ$ to reduce the probability of gate leakage in the large-area device required for ensemble measurements. The dopants were then activated by rapid thermal annealing at 900$^\circ$ for 10\:s in a nitrogen ambient. We found significant dopant pile up towards the dry oxide as measured by secondary ion mass spectroscopy, which shows an As concentration of $4-5\times10^{16}$\:/cm$^3$ in the top 50\:nm of the  $^{28}$Si epitaxial layer but drops to $\approx1\times10^{15}$\:/cm$^3$ at the $^{\rm nat}$Si-BOX interface, hence ~90\% of the implanted donors reside in the  $^{28}$Si layer. 

In order to access the deeply implanted $p+$ back gate in the handle wafer, contact windows were made by etching through the oxide layers and SOI layer. A BOSCH etch process was then carried out to etch 5\:$\mu$m down the handle wafer. After native oxide removal by hydrofluoric acid, Ti/Au metallic layers were subsequently  deposited to form both the contacts to the $p+$ back gate and defining the top gate electrode. The top gate electrode covers an area of $21\times2.6$\:mm$^2$ with approximately $2\times10^{11}$ donors underneath. A forming gas anneal (90\%\:$\rm N_2$, 10\%\:$\rm H_2$) at 350$^\circ$C for 20 minutes was then carried out to passivate the silicon-silicon dioxide interfaces in the structure. A schematic and the cross section of the final device structure are shown in Fig.\:\ref{f:1}.

Capacitance-voltage ($CV$) measurements at cryogenic temperatures reveal a gate capacitance of $1.9\:$nF/cm$^2$, in good agreement with the geometry of the device with the top 5\:$\mu$m of the silicon handle wafer being insulating. The capacitance is essentially constant within the gate voltage ($\leq 2$\:V) and pulse speed ($10~\mu$s) used in this study. 
Frequency dependent $CV$ measurements indicate that hole accumulation at the handle wafer-BOX interface was only achieved at low frequencies ($< 1\:$kHz), and on some devices only at temperatures greater than the carrier freeze-out temperature. 
We speculate that this is due to residual lattice defects in the handle wafer with sufficient trap density to prevent holes from the $p+$ back gate migrating to the interface efficiently.
Hence, the electric field strength is calculated assuming a metallic parallel-plate capacitor geometry, with dimensions as shown in the inset of Fig.\:\ref{f:1}.

All ESR measurements were carried out with a pulsed X-band ($\approx9.8$GHz and $B\approx 0.34$\:T) ESR spectrometer (Bruker Elexsys 580). A continuous wave ESR linewidth of $0.01$\:mT (0.3\:MHz) for the As donor was measured at 20\:K, a factor of $\sim$40 narrower than the known linewidth in $^{\rm nat}$Si.
In addition, an electron spin coherence time $T_2$=\:0.3\:ms was observed at 2\:K, indicating that the spin coherence is limited by the relatively high As concentration instead of $^{\rm nat}$Si in the SOI layer\:\cite{tyryshkin12}.
Both of these observations confirm that the signal mostly arises from dopants residing in the $^{28}$Si layer and that the SOI layer has a low defect density. 
All measurements discussed hereafter were performed at 15\:K, unless stated otherwise. 

Our parallel plate structure permits a uniform electric field to be readily applied to the As donors. Under small applied voltages the donor remains neutral at low temperatures, however, the electronic wavefunctions are distorted, causing a change in the contact hyperfine interaction and spin-orbit coupling strengths, hence inducing a Stark shift. We investigate the As Stark shift parameters using a Hahn echo pulse sequence, and we detect the frequency shift of the donors by examining the change in the Hahn echo signal phase as the applied voltage (electric field) is varied~\cite{bradbury06}.  In all cases, the static magnetic field $B_0$ is aligned along the $\langle 110\rangle$ axis, and orthogonal to the direction of the applied DC electric field $E$. 

We use a bipolar voltage pulse sequence (sequence IV in Ref [\cite{bradbury06}]), where voltages of opposite sign but equal amplitude are applied successively during the defocusing phase of the Hahn echo sequence. The synchronization of the microwave and voltage pulses is shown in Fig.~\ref{f:2}(a). The advantage of using a bipolar pulse sequence over a unipolar pulse sequence (containing only voltage pulses of the same sign) is that ambiguities arising from built-in electric fields cancel out. The donor resonant frequency shift $\Delta f_{m_I}$ for a given ESR line corresponding to nuclear spin $m_I$, due to the applied electric field of magnitude $|E|$ is given by:

\begin{equation}
	\label{e:1}
	\Delta f_{m_I}=\left(\frac{1}{h}\right)\left(\eta_aam_I+\eta_gg\mu_BB\right)E^2
\end{equation}

\noindent where $\eta_a$ is the hyperfine induced Stark parameter, $\eta_g$ the spin-orbit induced Stark parameter, $h$ Planck's constant and $\mu_B$ the Bohr magneton, while $a$ and $g$ are the hyperfine interaction strength and electronic $g$-factor, respectively. For As donors, $a=\:819$\:neV and $g=\:1.99837$~\cite{feher59}. We have neglected linear Stark shift terms ($\propto E$) which are much smaller compared with the quadratic effects in silicon\cite{bradbury06, rahman07}, and linear terms cancel out with the bipolar voltages pulses used. With such a frequency shift, a voltage (electric field) pulse of duration $\tau$ then induces a phase shift $\Delta\varphi_{m_I}=2\pi\Delta f_{m_I}\tau$ for the Hahn echo signal. 

All measurements were carried out with microwave pulse delays of $\tau_0$=\:15\:$\mu$s, and the voltage pulses for each polarity are 5\:$\mu$s long (total 10\:$\mu$s voltage pulse for each measurement).  A two step phase cycling (+X,+X and +X,$-$X) for the microwave pulses is used to remove spurious background signals. Figure~\ref{f:2}(b) shows the reference echo signal and Fig.~\ref{f:2}(c) shows the phase shifted echo signal under an applied electric field of $E=0.23$\:V/$\mu$m for the $m_I$=+3/2 line. The echo phase shifts extracted from the echo signals for the four different hyperfine-split lines are converted to frequency shifts and shown in Fig.~\ref{f:2}(d). The solid lines represent quadratic fits to the data as defined by Eq.\:\ref{e:1}. We find the contact hyperfine term  $\eta_a=\:-1.9\pm0.2\times10^{-3}\:\mu$m$^2$/V$^2$ and the spin-orbit term  $\eta_g=\:3\pm4\times10^{-6}\:\mu$m$^2$/V$^2$. The spin-orbit shift is very weak compared with the hyperfine shift under our experimental conditions with relatively moderate magnetic fields ~\cite{bradbury06, rahman09b}, and cannot be determined accurately within experimental errors. Indeed, fitting the data while neglecting the spin-orbit term results in virtually no change to the contact hyperfine value. We summarize the Stark shift parameters for various silicon shallow donors in Table~\ref{t:1}.  

\begin{table*}
	\centering
	\begin{tabular}{ c  c  c  c  l }
	\toprule
    	Donor & Orientation & $\eta_a$  & $\eta_g$ & Experiment/Theory \\ 
	 & & ($10^{-3}\mu$m$^2$/V$^2$) & ($10^{-6}\mu$m$^2$/V$^2$) & \\
	 \hline
	$^{31}$P & -- & $-2.8$ & -- & Tight-binding calculation.~\cite{rahman07}\\ 
	& $B\:||\:E$ & -- & -12 & Tight-binding calculation.~\cite{rahman09b}\\ 
	& $B\perp E$ & -- & +14 & Tight-binding calculation~\cite{rahman09b}\\ 
    	$^{75}$As & $B\perp E$ & $-1.9\pm0.2$ & $+3\pm4$ & ESR with parallel-plate geometry (present work). \\ 
    	$^{121}$Sb & $B\perp E$ & $-3.7$ & $-10$ & ESR with interdigitated electrodes.~\cite{bradbury06}\footnote{Note that the signs of the nuclear spin projections in Fig.\:4 of Ref\:[\cite{bradbury06}] were mislabeled, but the signs and values given for the Stark shift parameters are correct.} \\ 
	\botrule
	\end{tabular}
    \caption{\label{t:1}{Summary of the experimentally and theoretically determined Stark shift parameters for shallow donor in silicon.}}
\end{table*}

\begin{figure}
	\centering
	\includegraphics[width=8.0cm]{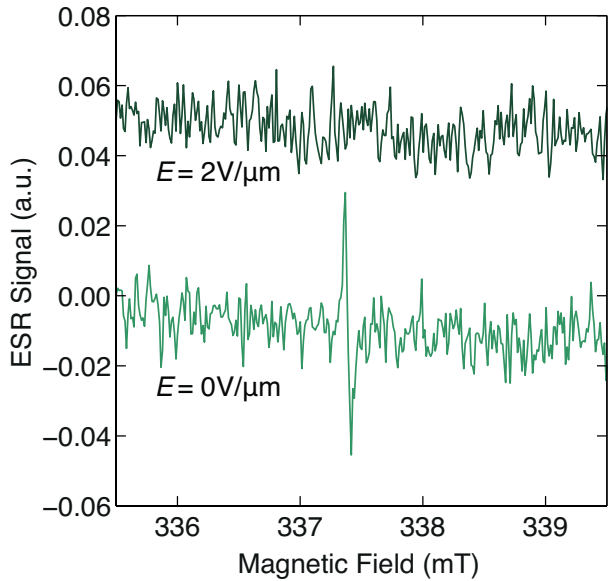}
	\caption{\label{f:3}{Continuous wave ESR signal of the $m_I$=+3/2 signal with constant voltage (lower) $E$=\:0V and (upper) $E=2\:\mu$m/V measured at 18\:K,  offset for clarity.}}
\end{figure}

We find that at larger electric fields the ESR signal intensity decays substantially due to the onset of dopant ionization, and that $\sim 2$V/$\mu$m is sufficient to completely field-ionize the donors, as shown in Fig.~\ref{f:3}. Removing the electric field causes the electron spin echo to reappear (i.e.\ re-neutralization of the donors) on a timescale of $\sim$1\:$\mu$s, limited by the RC constant of the device. The results of electric field-controlled dopant ionization on the nuclear spin coherence times will be reported in a subsequent publication.

The As dopant density used in this study corresponds to a mean donor-donor separation of approximately 40\:nm, which is close to the upper limit of exchange-based entanglement schemes for donor qubits. 
At these concentration levels, the spin coherence time is limited by direct dipole-dipole interaction amongst the donors~\cite{tyryshkin12}. We note that in a perfectly planer device architecture the dipole-dipole interaction can be removed by applying the magnetic field at the magic angle~\cite{desousa04}. However, the line widths are still dominated by inhomogeneous broadening, which is about 0.2\:MHz in the highest quality $^{28}$Si material, implying an electric field of the order of 1\:V/$\mu$m is needed for shifting an entire ESR line out of resonance as previously noted~\cite{bradbury06}. We found that even at 1\:V/$\mu$m the ESR signals are significantly reduced, indicating that some donors begin to be ionized by the moderate electric fields, which is consistent with theoretical calculations~\cite{calderon07} and electrical measurements~\cite{zurauskas84}. This underlines the difficulty of Stark tuning of donors relying on the hyperfine term alone. At higher magnetic fields when the spin-orbit term becomes appreciably larger~\cite{rahman09b}, or by inducing additional strain in the sample~\cite{dreher11}, larger Stark shifts can be induced to circumvent this issue.

In summary, we have investigated donor spin dynamics under the influence of applied electric fields with pulsed electron spin resonance. Arsenic donors are embedded in isotopically purified 28-silicon SOI parallel-plate structures formed between a metallic top gate and MeV implanted back gate. We measured the Stark shift of the donors by applying electric fields below the ionization threshold of the dopants, and find a hyperfine Stark shift parameter of  $\eta_a=-1.9\pm0.2\times10^{-3}\:\mu$m$^2$/V$^2$.

\acknowledgments
This research is supported by the EPSRC through the Materials World Network (EP/I035536/1) and CAESR (EP/ D048559/1) as well as by the European Research Council under the European Community's Seventh Framework Programme (FP7/2007-2013)/ERC grant agreement No.\ 279781. Work at Princeton was supported by NSF through Materials World Network (DMR-1107606) and through the Princeton MRSEC (DMR-0819860).  Work at LBNL was supported by the U.S. National Security Agency (100000080295) and by the U.S. Department of Energy (DE-AC02-05CH11231, LBNL). C.C.L. is supported by the Royal Commission for the Exhibition of 1851. S.S. is supported by the Violette and Samuel Glasstone Fund and St. JohnÕs College Oxford. J.J.L.M. is supported by the Royal Society. 
\bibliographystyle{unsrt}

\end{document}